\documentstyle[psfig,emulateapj]{article}

\lefthead{Tissera et al.}
\righthead{Chemical Enrichment at High Redshifts}

\begin{document}

\title{CHEMICAL ENRICHMENT AT HIGH REDSHIFTS:
UNDERSTANDING THE NATURE OF DAMPED Ly$\alpha$
SYSTEMS IN HIERARCHICAL MODELS}

\author{Patricia B. Tissera  \altaffilmark{1,4}}
\altaffiltext{1} {Instituto de Astronom\'{\i}a
y F\'{\i}sica del Espacio, Argentina
}
\author{Diego G. Lambas \altaffilmark{2,4}}
\altaffiltext{2} {Observatorio Astron\'omico 
de la Universidad Nacional de C\'ordoba,  Argentina
}

\author{Mirta B. Mosconi \altaffilmark{2}}

\author{Sofia Cora  \altaffilmark{3,4}}
\altaffiltext{3} {Observatorio Astron\'omico de
La Plata, Argentina
}

\altaffiltext{4}{ Consejo Nacional de Investigaciones Cient\'{\i}ficas
y T\'ecnicas
}

\begin{abstract}
We use cosmological hydrodynamical simulations including star formation 
and metal enrichment to study the evolution of the chemical properties of
galaxy-like objects at high redshift in the range  $0.25<z< 2.35$
 in a hierarchical clustering scenario.
As the galactic objects are assembled we find that their
gaseous components exhibit neutral Hydrogen column densities 
with abundances and scatter
comparable to those observed in damped Lyman-$\alpha$ systems (DLAs).
The unweighted mean of abundance ratios and 
 least
square linear regressions through the simulated DLAs
yield intrinsic  metallicity   evolution  
 for the [Zn/H] and [Fe/H],
consistent with results obtained from similar analysis of available
observations.  Our model statistically reproduces the mild evolution
detected in the metallicity of the neutral hydrogen content of the Universe,
given by mass-weighted means, 
if observational constraints are considered 
(as suggested by Boiss\'ee et al. 1998).
For the  $\alpha$-elements in the simulated DLAs,
 we find neither enhancement
 nor dependence
on metallicity. Our results support
the hypotheses  that DLAs trace a variety of galactic objects with different
formation histories and that both SNI and SNII are contributing 
to the chemical enrichment of the gas component at least since $z \approx 2$. 
This study indicates that DLAs could be understood as the
building blocks that merged to form today normal galaxies within a hierarchical
clustering scenario.

\end{abstract}


\section{INTRODUCTION}

Studies of DLA absorbers have provided hints on
the properties of structure at high redshift, from 
both  kinematical and chemical points of view (\cite{lu96}; \cite{pet97}; \cite{wol95};
Haehnelt, Steinmetz \& Rauch 1998; \cite{pet99}; \cite{pw99}).
Analysis of the absorbers at low $z$ show that these systems have
diverse morphologies (e.g., \cite{rao93}; \cite{brun97})
 while at high $z$ their nature remains unclear.
 First kinematical studies  suggested that
high redshift DLAs  were large disks similar to those in 
present-day spiral galaxies
(e.g., \cite{wol95}; \cite{pw97}). However, Haehnelt et al. (1998) show
that they are also  consistent with being the building blocks of local typical
galaxies,
as  predicted by hierarchical 
clustering scenarios (HCSs).
The metallicity  properties of DLAs support the idea that they are chemically
young objects
(e.g., \cite{pet99},  
 Prochaska et al. 2000).
However, the dependence of their metallicities on $z$ is a controversial
point since, until recently, only very weak  or non-evolution at all
has been reported from the analysis of the mean metallicity
of the neutral hydrogen with the abundance ratios weighted by the HI 
column density ($N_{\rm HI}$) 
(\cite{pw00}; \cite{pet99}; \cite{vla00}) in contrast
with some theoretical models which predict substantial evolution of
the neutral hydrogen abundances 
(e.g., \cite{ep97}; \cite{pei99}; see Cen \& Ostriker 1999 for a different
result). However, the unweighted mean metallicities 
and linear regressions through the individual data show stronger
evolution signals as reported in recent works (i.e., \cite{pw00};
Vladilo et al. 2000; Hou et al. 2001). Detecting evolution 
in chemical abundances of HI gas mass 
may be difficult  due to several biasing factors. Among them, dust depletion
and obscuration have been probed to be a source of uncertainties for
some element measurements (e.g., Vladilo 1998; \cite{hou01}), while the limited
redshift coverage with the known DLAs may not be adequate for detecting 
the presence of evolution. 
As first pointed out
by Boiss\'e et al. (1998), DLAs with
high ${ N_{\rm HI}}$
and high metallicity might be missing from the data as a consequence
of dust extinction
which would imply a biased determination of evolution.
For the $\alpha-$elements, dust-corrected data show mild enhancement
or near solar values
for most observed elements in contrast to the metallicity
pattern of metal-poor Galactic stars (e.g., \cite{vla98}; \cite{pet00}).
This might imply a different SF history than that of the Milky Way, although
 Prochaska  et al. (2000) found an agreement between the properties of
the Galactic thick disk and the DLAs.  

In HCSs, where galaxies are formed by aggregation of smaller substructures,
 mergers and continuous
gas infall play an important role,
contributing to regulate the star formation (SF; 
\cite{tis00}) and  chemical histories (e.g., \cite{cor00}) of 
the galactic objects. 
As suggested by Haehnelt et al. (1998) from a kinematical study of
numerical simulations, DLAs could be the progenitor
substructures of today galaxies  
 that merge to form galactic objects in
a hierarchical clustering scenario. 
The variety of morphological
types together with the fact that they tend to be
chemically young seem to support this hypothesis.
With the aim at understanding the possible link between the structure
in a HCS and the nature of DLAs, in this letter, we assess 
the chemical properties of 
today-galaxy building blocks 
as a function of the redshift.
For this purpose, we focus our study on the analysis of the unweighted 
mean metallicities of the
neutral hydrogen  
as a function of $z$ since they are more sensitive
to the properties of the individual objects. 
We also comment on the mass-weighted
mean metallicities that are related to the
chemical content of the Universe.
We use
hydrodynamical cosmological models that  treat the 
non-linear evolution of both baryons and dark matter in a self-consistent
way, providing a well-described history of formation, and include 
star formation and chemical evolution.
Results of the simulations are compared with available observational data
on chemical abundances of DLAs. We discuss  implications 
for galaxy formation.

\section{CHEMICAL PROPERTIES OF GALACTIC OBJECTS}

Our hydrodynamical chemical simulations follow the joint evolution of the
dark matter and baryons within a cosmological context (\cite{tis97})
including SF and chemical evolution.
Stars are formed from cold and dense gas in a convergent flow according
to the Schmidt law. Gaseous particles  are gradually
 transformed into stars in different SF episodes.
The contribution of type I (SNI) and type II (SNII)  supernovas 
from each of these SF episodes to the chemical
enrichment of the gas component are taken into account according
to stellar evolution models and metallicity enrichment yields.
Chemical elements generated in a given particle 
are distributed among gas particles within
its  neighboring area, weighting each contribution with a 
kernel function that depends on the distance (i.e., by
using the smooth particle hydrodynamics technique).
 
We adopt the yields given by Woosley \& Weaver (1995) for SNII
and those given by Thielemann, Nomoto \& Hashimoto (1993) for SNI.
 A time delay
of $10^8$ yrs is assumed for binary star systems to 
explode as SNI.  We adopt a fixed Salpeter Initial Mass
Function with lower and upper mass cut-offs of $0.1 \ {\rm M_{\sun}}$ 
and $120\ {\rm  M_{\sun}}$, respectively.
In this work the effects of energy injection into the
interstellar medium due to supernova explosions have not been included.
For a detailed discussion of the chemical model see Mosconi
et al. (2000).

We analyze cosmological simulations of a typical 10 Mpc cube volume
represented by $64^3$ equal mass particles ($\Omega_b=0.1$). 
Initial conditions are consistent
with a Standard Cold Dark Matter Universe
($H=50\ {\rm km \ s^{-1}Mpc^{-1}}$) with cluster abundance normalization,
 $\sigma_8=0.67$. 
We have run a set of three simulations with different realizations 
of the power spectrum, estimating averaged results over them.
The SN parameters adopted in these simulations correspond to  those  
giving the
best agreement with observations of galaxies at $z=0$ (\cite{mos00})
and the [O/Fe] abundance pattern in the Milky Way (\cite{cor00}).

We identified galactic objects at their virial radius at different stages of 
evolution of the simulated volume.
Galactic objects are formed by a dark matter halo and a baryonic component
in the form of gas and stars. 
To diminish numerical resolution problems,
we analyze  galactic objects  with more than 200 baryonic
particles within their virial radius and in the range $0.25 < z < 2.35$.
 Consequently,  the analyzed objects
have virial velocities within  
$\approx 100-250 \ {\rm km \ s^{-1}}$. 
We study a total number of 380 galactic
objects satisfying the above conditions which are all considered as possible
absorbers.

It is likely  that DLAs observations map the chemical properties
of the gaseous disks  
not expected to be  good tracers of the metal content at the central regions
 (e.g., \cite{jim99}; \cite{som01}; \cite{sav00}).
Hence, in order to  carry out a suitable comparison between the simulated
galactic objects and DLAs observations,
we  use a Monte Carlo technique to simulate 
line-of-sights (LOS) 
and estimate the chemical properties
of the neutral hydrogen component with  
 ${ N_{\rm HI}> 2 \times 10^{20} {\rm atoms/cm^{-2}}}$ (\cite{wol86}) along the LOS.
We assume that the hydrogen mass in a gas particle remains neutral if
no SF activity has ever occured within that particle.

As a combined result of dynamical evolution, mergers and interactions,
the SF rate history of each galactic  object
can be described as a contribution of an ambient SF rate and a
series of starbursts (e.g., \cite{tis00}).
The timings between starbursts are not ad hoc, but given naturally
by the evolution of the objects in the HCS adopted. Thus, we have 
a consistent description of the chemical 
enrichment of the stellar populations
and gaseous component due to the fact  that the different ejecta times of
SNI and SNII can be properly taken into account. 
Note that the simulated sample is not affected by dust depletion or
obscuration.
These facts make our simulations powerful tools
to explore the nature of DLAs.

For each galactic object selected at a given redshift we 
define the HI mass-weighted mean of abundance ratio of 
elements K and J,
[K/J] in the column of HI 
along a LOS, 
as:

\begin{equation}
[{\rm K/J}]={\rm log} \frac{\sum_{i=1}^{ n_{\rm p}} {\rm K}_i \,  M^{\rm gas}_i}
{\sum_{i=1}^{ n_{\rm p}} {\rm J}_i \,  M^{\rm gas}_i} - {\rm log}({\rm K/J})_{\odot},
\end{equation}

\noindent
where  $ n_{\rm p}$ is the total  number of gas particles belonging
to a galactic object 
along a certain  LOS,  
$M^{\rm gas}_i$ the neutral hydrogen mass of the $i^{th}$ 
particle,  ${\rm K}_i$ and ${\rm J}_i$ their chemical abundances, 
and $({\rm K/J})_{\odot}$ the corresponding solar abundance ratio.
Similarly, we can define the mean abundance ratios
for the stellar populations. This point will be discussed in a separate paper.
 
In order to study the metallicity evolution of DLAs, we consider
the abundance ratios of [Zn/H] and [Fe/H].
In our sample both elements can be measured without introducing biases  such
as those produced by dust extinction and depletion, or, in the case of zinc
by the fact that this element is difficult to detect at low metallicities.
Note that the zinc  is usually considered a reliable
tracer of metallicity because it is weakly depleted onto dust grains, 
 however, its nucleosynthesis 
remains to be fully understood (\cite{pw00}; \cite{pb00}). 
Current models for the production of this element (\cite{ww95}) have problem
in reproducing the observational fact that, regardless of metallicity, stars
in the halo and thin disk of the Galaxy have nearly 
solar [Zn/Fe] values (see also Hou et al. 2001).
 Nevertheless, recent results suggest that
Zn is enhanced relative to Fe  in the stars of the Galactic thick disk
(\cite{petal00})  These make results
from Zn analysis more difficult to interpret. 
For the purpose of  carrying out  a reliable comparison between observations and simulations,
we gathered the available observations from the literature (see Figure 1) and
estimated the same parameters calculated for the simulations within the
same redshift range. 

In Figure 1a, we show the unweighted mean 
${\rm <[Zn/H]>} = 1/n \sum_n{\rm [Zn/H]}$ ratios (where $n$ is the number of
galactic objects in the total simulated sample that are located at
a certain $z$)
for the HI component along the LOS of galactic objects in the
three simulations as a function of  redshift. 
Error bars correspond to the dispersion of the ratios of the galactic
objects at each analyzed $z$.
As it can be seen, the simulated neutral gas abundance ratios  
 are within the  observed range
for DLAs. In order to quantify their evolution with $z$,
  we
compute the unweighted mean
${\rm <[Zn/H]>}$
 in two redshift intervals, 
$z_{\rm low}=[0.26,1.5]$ and $z_{\rm inter}=(1.5, 2.35]$. 
We obtain ${\rm <[Zn/H]_{low}>}= -0.53 \pm 0.06 $ and 
${\rm <[Zn/H]_{inter}>}= -1.12 \pm 0.17 $ \footnote
{ Standard deviations
have been  calculated by using the 
resampling bootstrap technique with 500 random
samples.}.
These results imply an  evolution of $0.58 \pm 0.13$ dex between the low and
intermediate redshift bins in the models, while 
DLAs observations show a  variation of 
 $0.22 \pm 0.09$ for the same redshift bins.
The observed mean values  are ${\rm <[Zn/H]_{low}>}= -0.89 \pm 0.10 $
and ${\rm <[Zn/H]_{inter}>}= -1.11 \pm 0.07 $.
 Hence, 
at high $z$ the simulated
and observational values agree very well. However, the low-$z$  
values differ in $\approx 0.40$ dex. This difference 
in the lower $z$ interval could be attributed to   the observational bias
first discussed by Boiss\'e et al. (1998) that would prevent 
high $N_{\rm HI}$ systems with high metallicities 
to be detected (see also \cite{pb00};
Savaglio 2000).
By analyzing the chemical properties of the individual DLAs,
  Vladilo et al. (2000) found an anticorrelation signal
for the [Zn/H] data which yields  a linear regression with a slope of
$d {\rm log Zn}/dz=-0.32 \pm 0.13$, detecting intrinsic evolution for 
$z < 3.5$.  A similar analysis applied to our set of observational data
within our $z$-range  yields $d {\rm log Zn}/dz=-0.24 \pm 0.11$. 
For the simulated DLAs, we obtained a slope of 
$d {\rm log Zn }/dz= -0.42\pm 0.08$
which, albeit larger, is statistically consistent with observations.

Figure 1b shows  the unweighted 
${\rm <[Fe/H]>}$ versus $z$ for the same galactic 
objects plotted in Figure 1a. This element  suffers strong
gas depletion but 
 can be observed  over a larger $z$ interval, its nucleosynthesis
is better understood and can be easily detected at low metallicities.
>From Figure1b we see  that the estimated ratios for the  HI
along the LOS 
in the simulations match the  observational range
of  DLAs. Estimates 
of their  unweighted ${\rm <[Fe/H]>}$ 
at $z_{\rm low}$ and $z_{\rm inter}$ give
a 0.67 $\pm$ 0.13 dex evolution in the simulations.
Calculations for the available DLAs observations  in the range $ 0.25 < z
< 2.35$  show,
in this case,  a 
signal for evolution of $0.48 \pm 0.13$ dex. A linear regression
through the observed [Fe/H] data within the same $z$-range 
shows a slope of $-0.50 \pm 0.17$
(see also Savaglio 2000; Hou et al. 2001).
The linear correlation of the  simulated [Fe/H] ratios gives 
$d {\rm log Fe}/dz=-0.50 \pm 0.09$. All these values are in 
suitable
 agreement within the statistical uncertainties. 

The mass-weighted mean metallicities of HI  
 assess the evolution of the 
chemical properties of the
population as a whole and, consequently, of the chemical history 
of the Universe. As previously reported, both the observed
$N_{\rm HI}$-weighted mean
[Zn/H] and [Fe/H] ratios show mild or
no evolution for $z \le 3.5$ (\cite{pet99}; \cite{pw00}),
while most theoretical models predict substantial
change of the metallicity of neutral gas with redshift. Estimates
of the corresponding ratios in our model at low and intermediate $z$ 
intervals also yield 
a signal of 
evolution: $0.33 \pm 0.08$ and $0.47 \pm 0.09$ for the [Zn/H] and [Fe/H],
respectively. 

We would like to stress that the conclusions about the evolution of the
observed [Zn/H] and [Fe/H] by using mean values at low and intermediate $z$
should be taken with caution because of
the small number statistics in the low $z$-bin. More observations
of DLAs at low $z$ are needed to draw final conclusion on this point.  
But, mainly, more observations of DLAs with high ${ N_{\rm HI}}$  and
high metallicity are required to properly assess the presence of
evolution. 
In fact, if DLAs with these characteristics were missing from the data then
the presence of evolution could be difficult to detect.
In order to mimic this effect, Prantzos \& Boissier (2000) applied a filter to their model by selecting
the  DLAs according to 
the observationally determined constraint
$18.8 < {\rm [Zn/H] + log} N_{\rm HI} < 21$. 
 Following their work,
we estimate the evolution of the filtered simulated DLAs,
using the [Zn/H] ratios.
For the unweighted means we obtained an evolution signal of $0.46 \pm 0.10$
while the weighted ratios show a weaker change consistent
with mild evolution: $0.24 \pm 0.12$.
Clearly, selection biases could have strong implications when estimating
changes with redshift.   

The standard 
dispersions shown in Figure 1 for both elements in the simulated sample
reproduce fairly well the observed  dispersions of DLAs: for the [Zn/H]
the building blocks show a dispersion of $\sigma_{\rm z_{low}}=0.68$
and $\sigma_{\rm z_{inter}}=0.78$  while observations have 
$\sigma_{\rm z_{low}}=0.42$
and $\sigma_{\rm z_{inter}}=0.86$.  The simulated [Fe/H] ratios have
 $\sigma_{\rm z_{low}}=0.60$ and  $\sigma_{\rm z_{inter}}=0.77$ 
compared to the observed ones of  $\sigma_{\rm z_{low}}=0.45$ and
 $\sigma_{\rm z_{inter}}=1.24$. In our models 
this dispersion arises as the results of  the variety of  galactic objects
 with diverse 
evolutionary histories: merger tree, star formation, chemical enrichment, etc.,
at different $z$, supporting a similar origin for the large dispersion   observed in DLAs.

We estimate the  unweighted mean [Si/Fe] and [S/Fe] ratios
in order to analyze the
behavior of the $\alpha$-elements in the HI gas mass of 
the simulated galactic objects. 
>From Figure 2a we can appreciate that the mean [Si/Fe] ratios
have nearly  solar values and show no dependence on metallicity,
consistent with the dust-corrected
observations reported by Vladilo (1998) and Pettini et al. (2000)
who assumed that Zn is undepleted and traces Fe.
A similar behavior was found for mean [S/Fe] ratios.

In Figure 2b we show the  unweighted ${\rm <[Zn/Fe]>}$ abundances
versus ${\rm <[Fe/H]>}$ .
We find that the HI component of the simulated
DLAs tend to exhibit  [Zn/Fe] values
around a mean of $0.27 \pm 0.28$ (1$\sigma$ deviation).
 It should be  noted that these
values are  consistent
with the observed range for DLAs with no dust corrections 
(with a mean of $0.47 \pm 0.31$) while dust-corrected ratios show nearly 
solar values.
As more extensively discussed by Prantzos \& Boissier (2000; see 
also \cite{lu96} and \cite{pw00})  current  
nucleosynthesis models of the zinc have
 problem in reproducing the observed abundance pattern of  the Milky-Way.
If we had assumed that Zn traces Fe for all metallicites 
(e.g., \cite{pet00}), 
our models 
would have produced  mean ratios in agreement with those of Vladilo (1998)
corrected by dust-depletion.

\section{DISCUSSION AND CONCLUSIONS}

Observational studies of the chemical properties
 of DLAs have provided unvaluable
information on galaxy evolution up to $z\approx 4$. 
This letter aims to provide a theoretical 
frame to understand these 
observations within  HCSs. The chemical hydrodynamical cosmological
models analyzed take into account different  physical
mechanisms controlling the formation and evolution of
galaxies:
gravitational collapse, tidal torques,
mergers, interactions, gas infall, gradient pressure,
radiative cooling, SF and chemical enrichment by SNI and SNII explosions.
In this scenario we follow the formation and evolution of galaxy-like
objects, analyzing the chemical properties of the different
substructures that merge hierarchically.

In order to properly compare the galactic objects in the simulations
with the observed DLAs, we estimated the abundances of HI
along LOS randomly distributed.
We found that the simulated galactic objects have gaseous components  that,
when randomly sampled, have 
[Zn/H] and [Fe/H] abundances within the observed range for DLAs. 
In the simulations, the unweighted [Zn/H]  and [Fe/H] ratios show 
evolution with $z$ in statistical agreement with results obtained
from similar analysis of the available  observations  
(see also Vladilo et al. 2000; \cite{pw00}; Hou et al. 2001).
However, the mass-weighted metallicities of the simulated and observed
$ N_{\rm HI}$ show
opposite trends; while the former show  evolution, the later do not.
 Part of this discrepancy
can be attributed to observational biases that might prevent
low-density metal-poor and very metal-rich and dense zones to be
detected by current observations. In fact, 
taking into account these effects
by applying a filter
to the simulated [Zn/H],
we found, on one hand,  that the unweighted means
still show a clear evolution, and, on the other,
that  the  
filtered weighted abundances yield a weaker dependence with $z$  
in agreement with corresponding observations (e.g., Pettini et al. 1999). 
Accordingly to our results, 
this observational bias seems to affect more strongly the weighted
means suggesting that the interpretation of the evolution of the chemical
content of the Universe by using DLAs should be taken with extra caution.
More observations, principally of high density and high
metallicity HI column densities, are needed to draw a final conclusion.

The lack of $\alpha-$enhancements in the {\it mean} abundances of 
simulated galactic objects
 shows that, in the redshift range analyzed, there are already significant
contributions 
from both types of SNs, so that the mean unweighted metallicities of the
$N_{\rm HI}$ 
are nearly solar, in agreement with DLAs observations
after dust corrections. 
We detect no evolution of [$\alpha$/Fe]  with metallicity
implying that, for our galactic objects, their SF histories
are the result of the superposition of different starbursts occuring
at different epochs. Moreover, the starbursts could have been triggered
in different substructures that merge to form the galactic objects
analyzed at a given z.
In the case of Zn we found 
 enhancement  with respect to the Fe but no dependence on metallicity.
A lower Zn nucleosynthesis production  in the models (or assuming
that Zn traces Fe) would have produced [Zn/Fe] values in agreement with the 
dust-corrected observed DLAs (\cite{vla98}). This lower Zn production
does not affect the agreement found between the observed and simulated
[Zn/H] at low $z$ shown in Figure 1a. At high z, models and observations
can be reconciled  if one takes into account that several high $z$ estimations
of [Zn/H] in observed DLAs are  upper limits instead of
accurate values.
The zinc remains a controversial element as pointed out by several
authors (\cite{lu96}; \cite{pet99}; \cite{pet00}; \cite{pb00}; \cite{sav00})
whose production and evolution are not fully understood.

To sum up, the chemical properties of galactic objects in our models at 
a given z are the result of their past evolution: 
collapse, mergers, interactions, SF history, etc. Since
we do not impose any particular constraint to select
the galactic objects in the models, the  results of our chemical
analysis  seem to support 
that DLAs could be the building blocks  of today normal galaxies within
a hierarchical context. 
\acknowledgments

We thank the anonymous Referee for a careful reading and thoughtful
comments that helped to improve this letter.
This work was partially supported by the 
 Consejo Nacional de Investigaciones Cient\'{\i}ficas y T\'ecnicas,
Agencia de Promoci\'on de Ciencia y Tecnolog\'{\i}a,  Fundaci\'on Antorchas
 and Secretaria de Ciencia y 
T\'ecnica de la Universidad Nacional de C\'ordoba.
P. Tissera thanks the hospitality of Observatorio Astron\'omico de C\'ordoba
during her visits.


\clearpage

\begin{figure}
\caption{[Zn/H] (a) and [Fe/H] (b) unweighted mean abundances for 
the neutral hydrogen along LOS  
in the galactic
objects in the three simulations as a function of the redshift
({\em filled circles}).  Error bars correspond to $1\sigma$ standard
deviation.
We include observational data ({\em open pentagons}) for DLAs taken from
Lu et al. (1996), Pettini et al. (1997; 1999; 2000), Vladilo (1998)
and Prochaska \& Wolfe (1999; 2000). 
The downward- (upward-) pointing arrows indicate upper (lower) limits of
[Zn/H].
Unweighted mean values for 
the simulated DLAs  ({\em dotted-open circles})
and the observations ({\em dotted-open pentagons}) in the low and intermediate 
redshift intervals are also plotted. These symbols are
superimposed 
in the intermediate redshift interval in Figure (a).
Solid lines are the least square linear regression for  
 the whole sample of simulated DLAs.
}
\end{figure}

\begin{figure}
\caption{[Si/Fe] (a) and [Zn/Fe] (b)  unweighted mean abundances 
for neutral hydrogen along LOS  in the simulated DLAs  
in the three simulations
for each redshift analyzed
as a function of the metallicity [Fe/H] 
({\em filled circles}). Error bars correspond to $1\sigma$ standard deviation. 
We include observational
data for DLAs without dust corrections ({\em open pentagons})
taken from Lu et al. (1996) and Prochaska \& Wolfe (1999); dust corrected
values ({\em dotted-open pentagons}) are obtained from Vladilo (1998) and 
Pettini et al. (2000).
The downward- (upward-) pointing arrows indicate upper (lower) limits of
[Si/Fe] and [Zn/Fe].
}
\end{figure}


\begin{thebibliography}{}


\bibitem[Boiss\'e et al. 1998]{bos98} Boiss\'e, P., Le Brun, V., Bergeron, J.,
\& Deharverng, J. M. 1998, A\&A, 333, 841

\bibitem[Cen \& Ostriker 1999]{cen99}Cen, R., \& Ostriker, J. P. 1999, ApJ, 519, L109

\bibitem[Cora et al. 2000]{cor00}Cora, S. A., Mosconi, M. B., Tissera, P. B., \&
Lambas, D. G. 2000, Proceeding of Stars, Gas and Dust in Galaxies: Exploring the Links
(astro-ph/0007072)

\bibitem[Edmunds \& Phillips 1997]{ep97}Edmunds, M. G. \& Phillips, S. 1997, MNRAS, 292, 733


\bibitem[Haehnelt et al. 1998]{haeh98}Haehnelt, M . G., Steinmetz, M., \& Rauch, M. 1998, \apj, 495, 647

\bibitem[Hou et al. 2001]{hou01}Hou, J. L., Boissier, S., \& Prantzos, N. 2001,
A\&A, in press (astro-ph/0102188)

\bibitem[Jimenez et al. 1999]{jim99}Jimenez, R., Bowen, D. V., \& Matteucci, F. 1999,
ApJ, 514, L83

\bibitem[Le Brun et al. 1997]{brun97} Le Brun,  V., Bergeron, J., Boiss\'e, P.,
\& Deharveng, J. M. 1997, A\&A, 279, 733

\bibitem[Lu et al. 1996]{lu96}Lu, L., Sargent, W. L. W.,  Barlow, T. A.,
Churchill, C. W., \& Vogt, S. S.  1996, ApJS, 107, 475

\bibitem[Mosconi et al. 2000]{mos00}Mosconi, M. B., Tissera, P. B., Lambas, D. G.,
\& Cora, S. A. 2000, \mnras, in press (astro-ph/0007074)

\bibitem[Pei et al. 1999]{pei99}Pei, Y. C., Fall, S. M., \& Hauser, M. G. 1999, ApJ, 522, 604

\bibitem[Pettini et al. 1997]{pet97} Pettini, M., Smith, L., King, D.,
\& Hunstead, R. 1997, ApJ, 486, 665

\bibitem[Pettini et al. 1999]{pet99}Pettini, M., Ellison, S. L., Steidel, C. C.,
\& Bowen, D. V. 1999, ApJ, 510, 576

\bibitem[Pettini et al. 2000]{pet00}Pettini, M., Ellison, S. L., Steidel, C. C.,
Shapley, A. E., \& Bowen, D. V. 2000, ApJ, 532, 65

\bibitem[Prantzos \& Bossier 2000]{pb00}Prantzos, N., \& Boissier, S. 2000,
MNRAS, 315, 82

\bibitem[Prochaska \& Wolfe 1997]{pw97}Prochaska, J. X., \& Wolfe, A. M. 1997, ApJ, 487, 73

\bibitem[Prochaska \& Wolfe 1999]{pw99}Prochaska, J. X., \& Wolfe, A. M. 1999, ApJS,
121, 369

\bibitem[Prochaska \& Wolfe 2000]{pw00}Prochaska, J. X., \& Wolfe, A. M. 2000, ApJ, 533, L5

\bibitem[Prochaska et al. 2000]{petal00}Prochaska, J. X., Naumov, S. O.,
Carney, B. W., McWilliam, A., \&
Wolfe, A. M. 2000, AJ, 12, 2513

\bibitem[Rao \& Briggs 1993]{rao93}Rao, S. M., \& Briggs, F. H. 1993, ApJ, 419, 515

\bibitem[Savaglio 2000]{sav00}
 Savaglio, S. 2000, Harwit, M.,
Hauser, M., eds., The Extragalactic Infrared Background and its Cosmological
Implications, IAU, vol. 204 (astro-ph/0011473)

\bibitem[Somerville et al. 2001]{som01}Somerville, R. S., Primack, J. R.,
\& Faber, S. M. 2001, MNRAS, 320, 504.

\bibitem[Thielemann et al. 1993]{thi93}Thielemann, F. K., Nomoto, K., \& Hashimoto, M.  1993,
  Prantzos, N., Vangoni-Flam, E., Cass\'e N., eds.,
Origin and Evolution of the Elements,
 p.299

\bibitem[Tissera 2000]{tis00}Tissera, P. B. 2000, ApJ, 534, 636

\bibitem[Tissera et al. 1997]{tis97} Tissera, P.B., Lambas, D.G., \& Abadi, M.G.
1997, MNRAS, 286, 384

\bibitem[Vladilo 1998]{vla98}Vladilo, G. 1998, ApJ, 493, 583

\bibitem[Vladilo et al. 2000]{vla00} Vladilo, G., Bonifacio, P.,
Centurion, M., \&
Molaro, P. 2000, ApJ, 543, 24 

\bibitem[Wolfe et al. 1986]{wol86}Wolfe, A. M., Turnshek, D. A., Smith, H. E.,
Cohen, R. D., 1986, ApJS, 61, 249

\bibitem[Wolfe et al. 1995]{wol95}Wolfe, A. M., Lanzetta, K. M., Foltz,
C. B., \&
Chaffee, F., J. 1995, ApJ, 454, 698

\bibitem[Woosley \& Weaver 1995]{ww95}Woosley, S. E., \& Weaver, T. A. 1995, ApJS, 101, 181

\end{thebibliography}
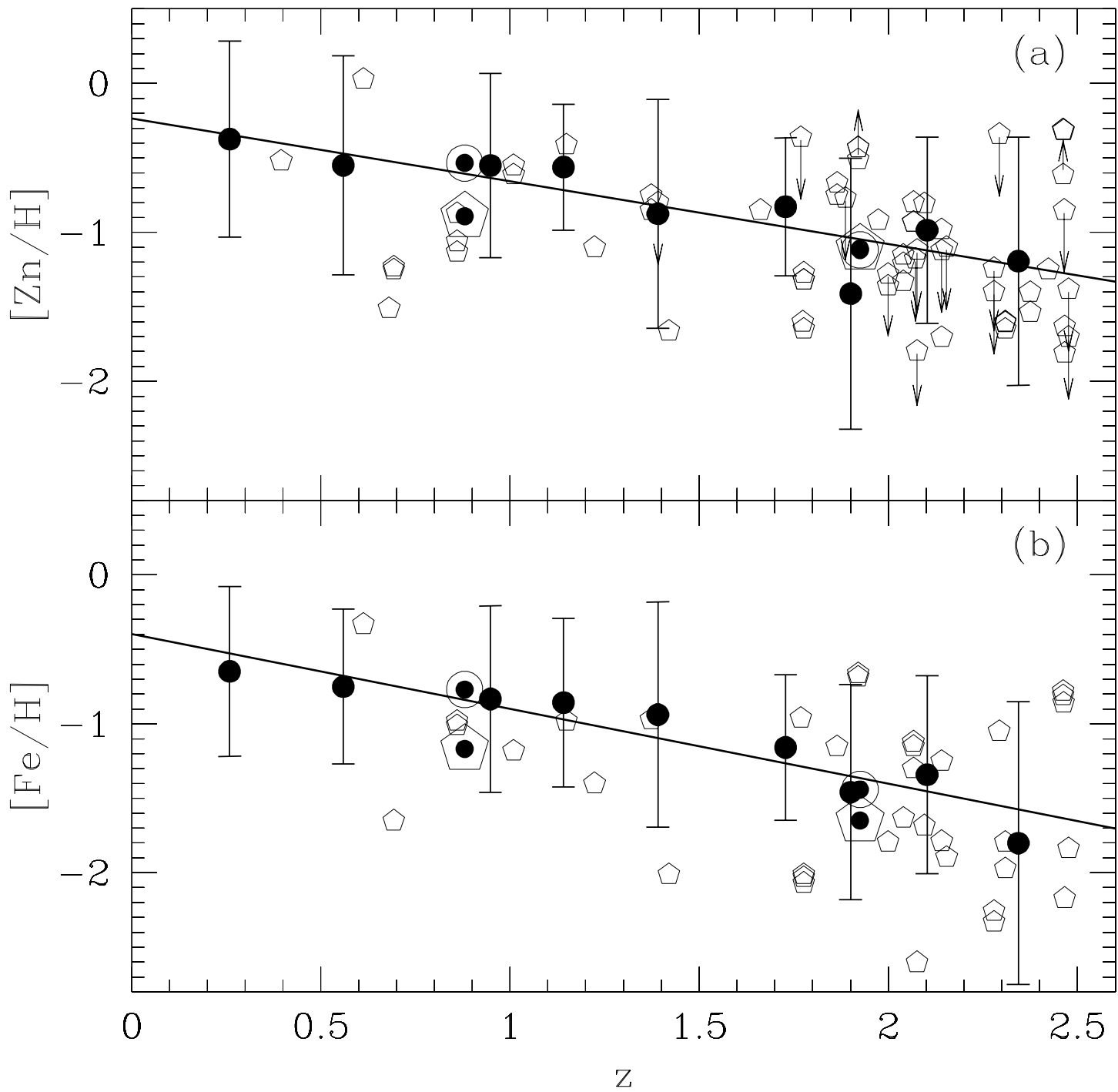
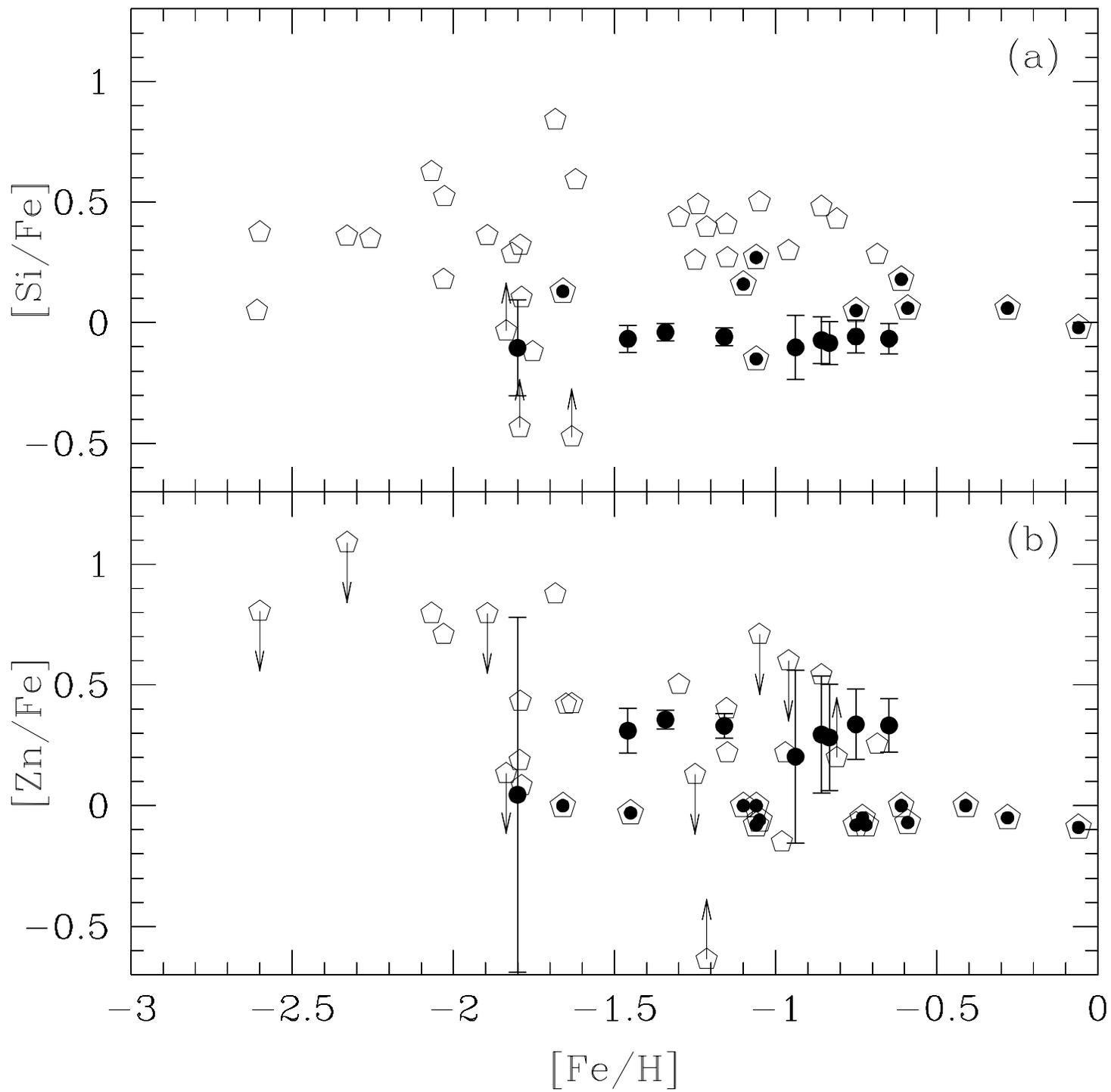
\end{document}